\documentclass[3p,times]{elsarticle}

\usepackage{ecrc}

\usepackage{epstopdf}

\volume{00}

\firstpage{1}

\journalname{Nuclear Physics A}

\runauth{R. Belmont}

\jid{nupha}

\jnltitlelogo{Nuclear Physics A}

\usepackage{graphicx}
\usepackage{amsmath,amssymb}

\begin{document}

\begin{frontmatter}



\title{Charge-dependent anisotropic flow studies and the search for the Chiral Magnetic Wave in ALICE}

\author{R. Belmont (for the ALICE\fnref{col1} Collaboration)}
\fntext[col1] {A list of members of the ALICE Collaboration and acknowledgements can be found at the end of this issue.}
\address{Wayne State University}




\begin{abstract}

Theoretical calculations have shown the possibility of P-violating bubbles in the QCD
vacuum, which in combination with the strong magnetic field created in off-central
heavy-ion collisions lead to novel effects such as the Chiral Magnetic Effect (CME)
and the Chiral Separation Effect (CSE).
A coupling between the CME and the CSE
produces a
wave-like excitation called the Chiral Magnetic Wave (CMW).  The CMW produces a
quadrupole moment that always has the same sign and is therefore present in an average
over events.  In this talk we present a series of charge-dependent anisotropic
flow measurements in Pb--Pb collisions at $\sqrt{s_{NN}}$ = 2.76 TeV in ALICE,
using two- and three-particle correlators with unidentified hadrons.
The relation of these measurements to the
search for the CMW is discussed.

\end{abstract}


\end{frontmatter}



\section{Introduction}
\label{intro}

Off-central heavy-ion collisions create an almond-shaped overlap region, which extends
above and below the reaction plane.  The pressure gradients in-plane are greater than
those out-of-plane, which creates an azimuthal anisotropy in momentum space as the system
expands.  Additionally, the spectator protons can be thought of as small but very dense
currents.  Since there are two currents close-by pointing in opposite directions, the induced
magnetic field from each current adds linearly in the region between them, creating a very
large and relatively homogeneous magnetic field in the same location in configuration space
as the medium created in the overlap region.  The interaction of the magnetic field with 
the produced particles in a region of space with topologically non-trivial gluon field
configurations leads to novel effects like the Chiral Magnetic Effect
(CME)~\cite{refCME} and the Chiral Separation Effect (CSE)~\cite{refCSE}.

The CME and CSE can be summarized succinctly in a pair of equations~\cite{refCMWa},
\begin{eqnarray}
\vec{J}_V &=& \frac{N_c e}{2\pi^2}\mu_A\vec{B}, \label{eq:cme}\\
\vec{J}_A &=& \frac{N_c e}{2\pi^2}\mu_V\vec{B}. \label{eq:cse}
\end{eqnarray}
Equation~\ref{eq:cme} is for the CME, and indicates that a vector current
$J_V$ (for example an electric current) is coupled to an axial chemical potential
$\mu_A$, oriented along the the magnetic field $B$.
Equation~\ref{eq:cse} is for the CSE, and indicates that an axial current
$J_A$ is coupled to a vector chemical potential $\mu_V$
(for example the scalar electric potential),
again oriented along the the magnetic field.

\begin{figure}[h!]
\begin{center}
\includegraphics*[width=9.0cm]{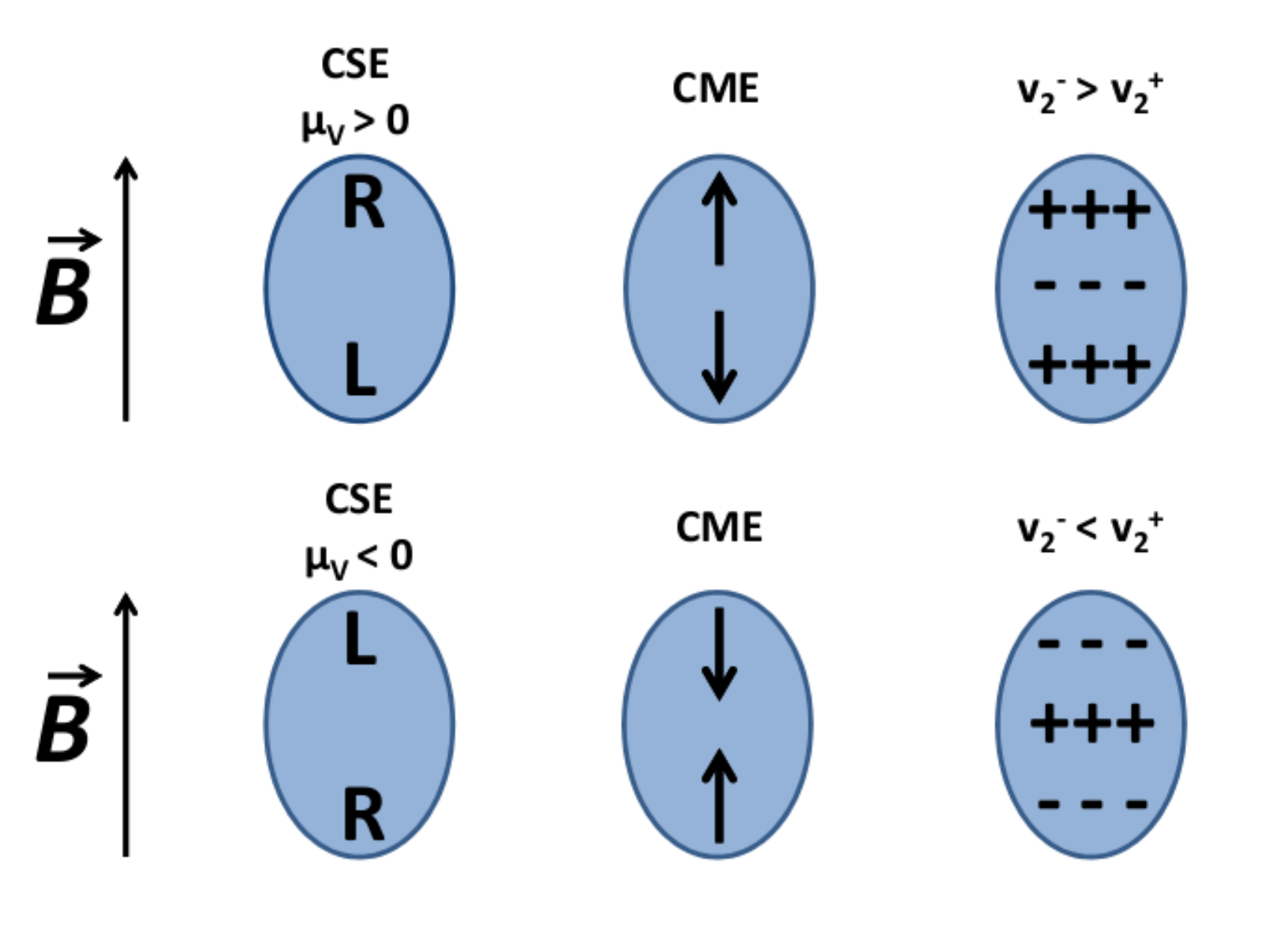}
\caption{Cartoon showing the basic picture of the
Chiral Magnetic Wave for the two cases $\mu_V>0$ and $\mu_V<0$.}
\label{fig:cartoon}
\end{center}
\end{figure}

One can tell by inspection that the two currents are coupled.  By changing to the
chiral basis, $V=R+L$ and $A=R-L$, one can derive two equations indicating
two electric currents that always point in opposite directions, leading to an electric
quadrupole moment that always has the same sign.  A detailed explanation and derivation
is given in~\cite{refCMWa}.
This effect is shown schematically in
Figure~\ref{fig:cartoon}.


In Figure~\ref{fig:cartoon}, the upper row shows an example where the plasma has
a positive electric charge state, i.e. $\mu_V>0$.  This causes a CSE current pointing
upwards, leading to an excess of
right-handed particles above the reaction plane and an excess of left handed
particles below it.  This means that above the reaction plane one has $\mu_A>0$
and below it one has $\mu_A<0$.  This then leads to two oppositely directed
CME currents, each pointed away from the reaction plane.  This leads to a positive
electric quadrupole, with excess positive charges out of plane at the poles and
excess negative charges in plane at the equatorial region.  Finally, under the
hydrodynamic expansion of the medium, the equatorial region has a larger flow
velocity due to the larger pressure gradients in plane, and therefore one observes
large $v_2$ for negative particles than positive particles, i.e. $v_2^->v_2^+$.
The lower row of the figure shows the same schematic for the opposite case with
$\mu_V<0$, which leads to exactly the same effect with all signs flipped.
A detailed explanation and derivation of how the presence of CMW affects the
final state observables in this way is given in~\cite{refCMWb}

\section{Methodology and observables}

From the previous section, the most intuitive observable would be $v_2$
(or $v_n$ to explore possible higher harmonic effects) as a function of
the event charge asymmetry $A$, which is defined as
$A = \frac{N^+-N^-}{N^++N^-}$,
where $N^+$ and $N^-$ represent the number of positive and negative particles,
respectively, measured in some well-defined region of phase space, i.e. for
some $p_T$ selection and, more importantly, some $\eta$ acceptance.
Indeed, this observable has been proposed theoretically~\cite{refCMWb} and
measured experimentally in Au--Au collisions at $\sqrt{s_{NN}}$ = 200 GeV
by the STAR collaboration~\cite{refSTAR}.  However, one of the issues with
this observable is that the slope of $v_n$ vs $A$ is not independent of
experimental effects (for example tracking efficiency) and therefore
requires a correction factor.

In this analysis, we use a novel three particle correlator~\cite{refSergei}
that is independent of efficiency and therefore requires no correction.
The three particle correlator is $\langle\cos(n(\phi_1-\phi_2))q_3\rangle$,
where $\phi_1$ and $\phi_2$ are the azimuthal angles of particles 1 and 2,
and $q_3$ is the charge ($\pm$1) of particle 3.  The
$\cos(n(\phi_1-\phi_2))$ part is estimated using the cumulant method and
denoted as $c_n\{2\}$.  In the absence of charge dependent correlations,
the correlator should factorize, i.e.
\begin{equation}
\langle \cos(n(\phi_1-\phi_2)) q_3 \rangle - \langle q_3 \rangle \langle \cos(n(\phi_1-\phi_2)) \rangle = 0.
\end{equation}
Note that when
the charge of the third particle is averaged over all particles in the event
(in the specified kinematic acceptance), the mean is equal to the charge
asymmetry, i.e. $\langle q_3\rangle = \langle A\rangle$.

\section{Results}

\begin{figure}[h!]
\begin{center}
\includegraphics*[width=8.0cm]{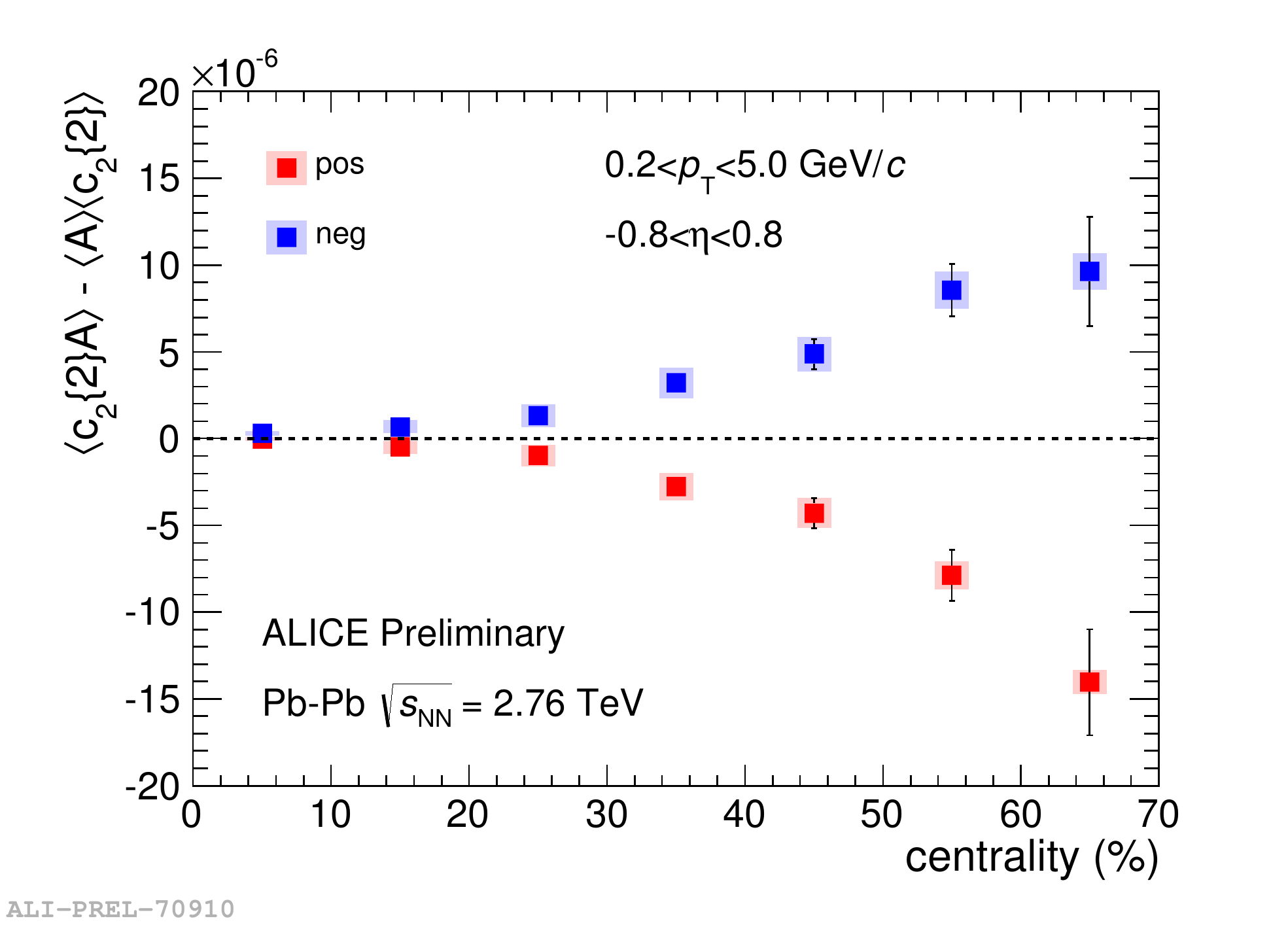}
\includegraphics*[width=8.0cm]{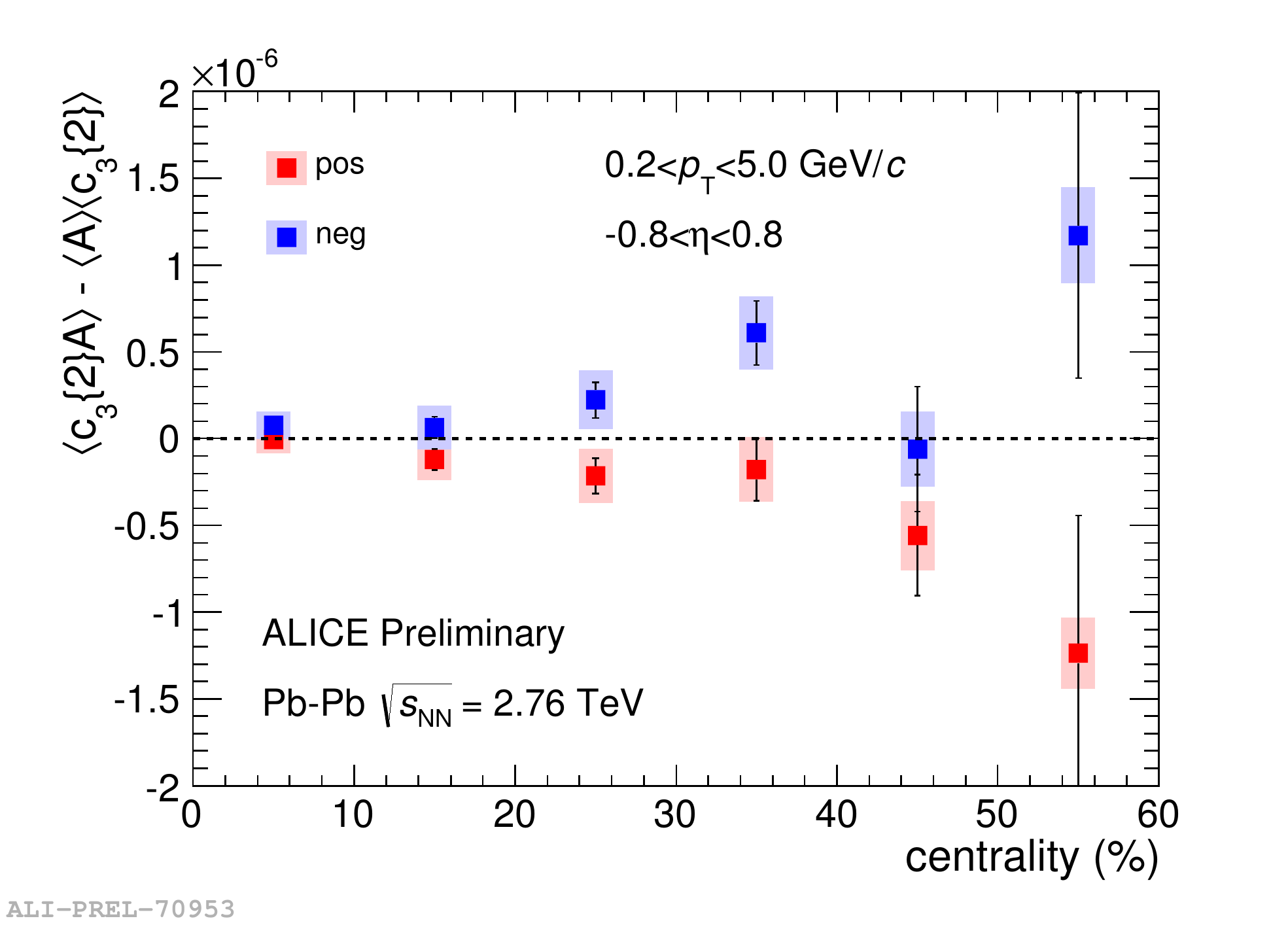}
\caption{Three particle correlator for the second (left panel) and third
(right panel) harmonic as a function of centrality.}
\label{fig:c232Avscent}
\end{center}
\end{figure}

Figure~\ref{fig:c232Avscent} shows the three-particle correlator for the second
(left panel) and third (right panel) harmonic as a function of centrality.
For the second harmonic one sees a substantial increase in the correlation
strength as the collisions become more peripheral.  This could be caused by
any possible combination of several factors.  The magnetic field strength
increases as the impact parameter increases
and thus the current gets stronger.  This would cause the correlations due
to the CMW to get stronger.  Additionally, effects due to local charge conservation
(LCC, i.e. the creation of balanced charge pairs close by in position space)
could play a role~\cite{refScott}.  Since central collisions have more
combinatoric (uncorrelated) pairs, the correlations due to LCC suffer combinatorial
dilution.  Note that neither of these necessarily comes at the expense of the other.
Non-flow 3-particle correlations may also contribute, for example from jet-like
correlations and 3-body decays.  However, a comparison to HIJING Monte Carlo was
performed and the three-particle correlator was found to be statistically
insignificant, indicating these effect play at most a very small role.
For the third harmonic one sees a small but non-zero correlation.  The physical
origin is not clear, though background from LCC is likely a contributor.
Additional factors could also contribute, such as interference between the second
and third harmonics, higher order multipole moments of the P-violating effects,
and other possible effects.

\begin{figure}[h!]
\begin{center}
\includegraphics*[width=8.0cm]{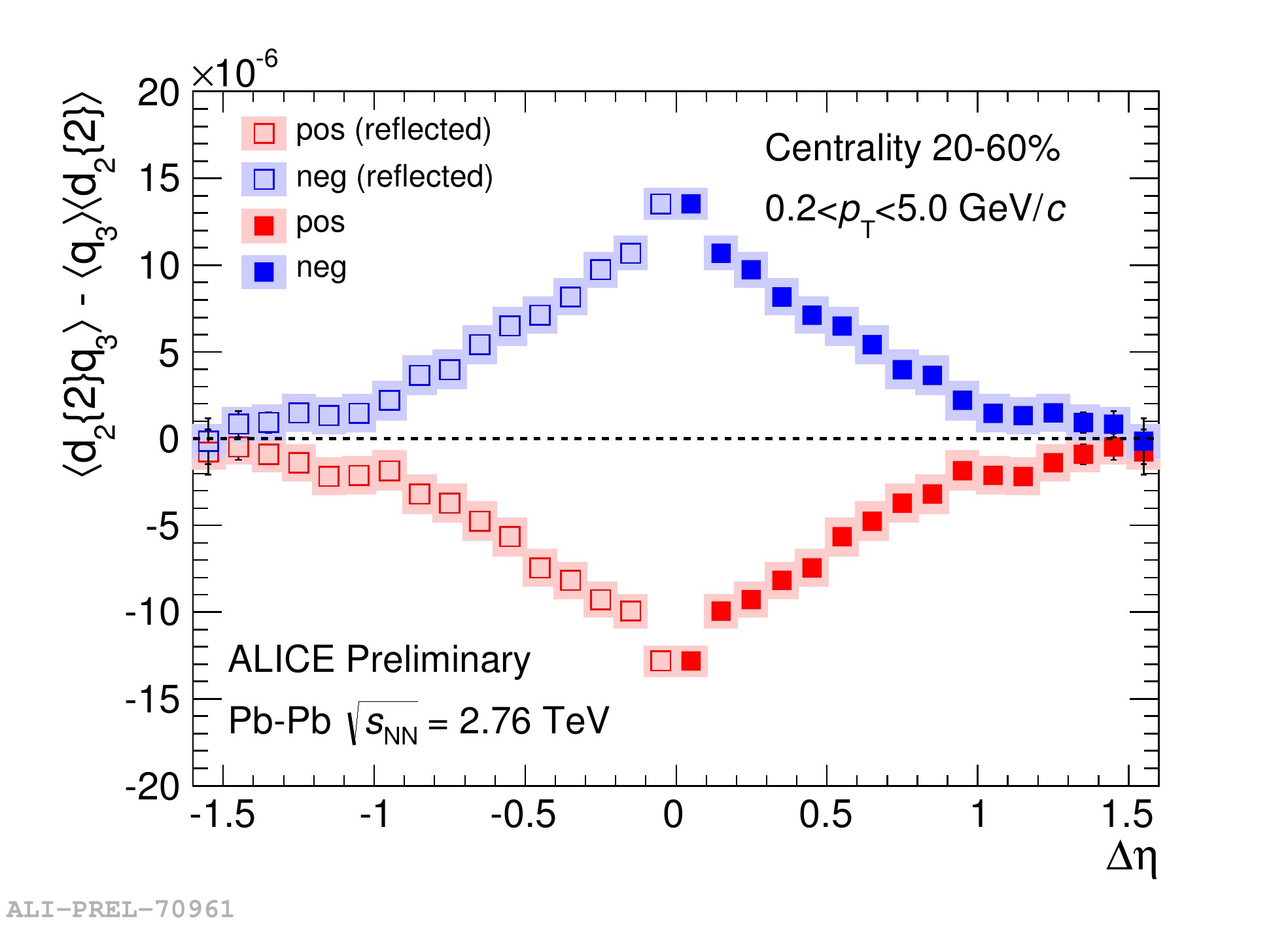}
\includegraphics*[width=8.0cm]{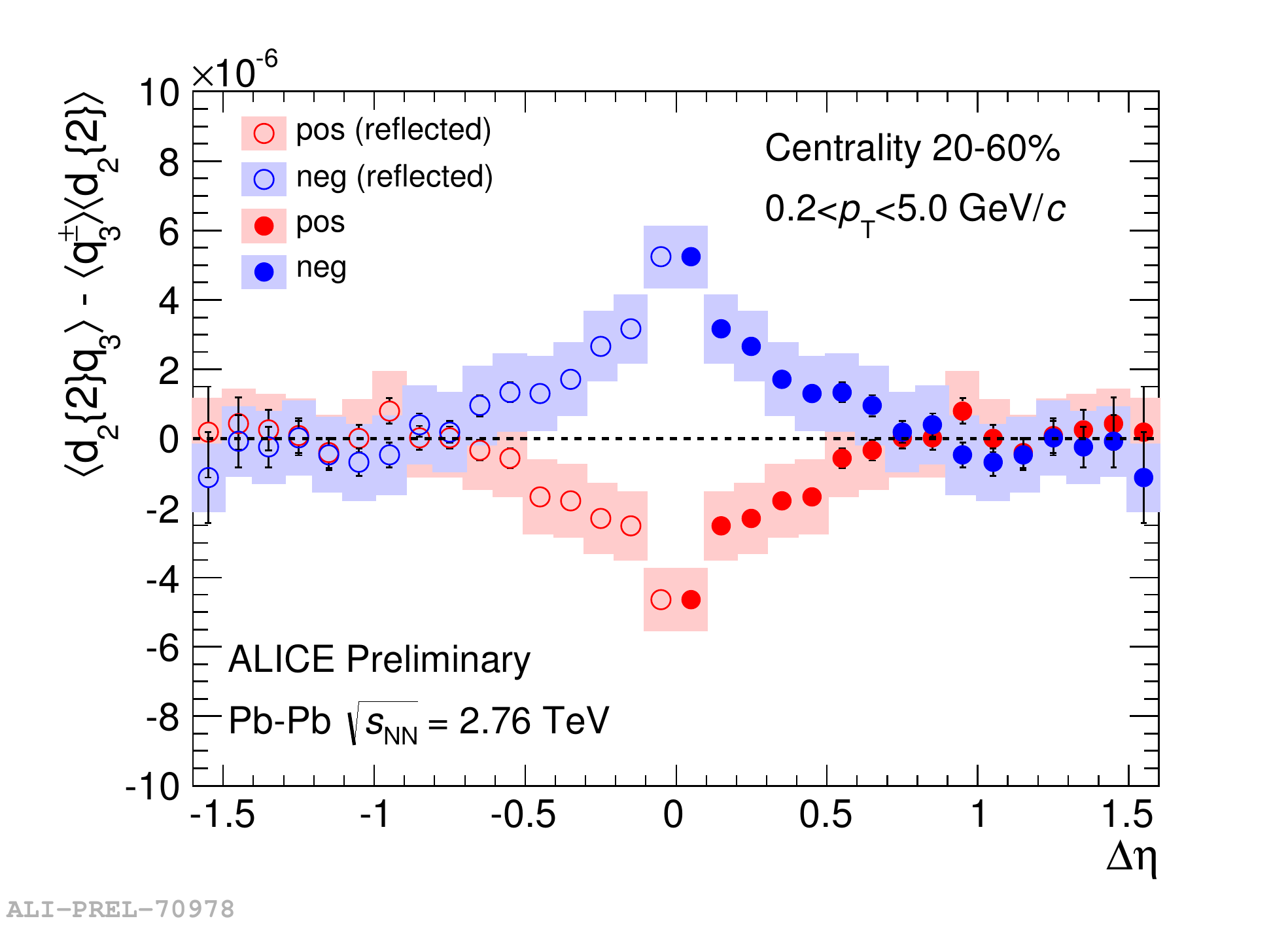}
\caption{Three particle correlator for the second harmonic.
The left panel shows the results with charge independent subtraction
and the right panel shows the results with charge dependent subtraction.}
\label{fig:c22Avsdeta}
\end{center}
\end{figure}

Figure~\ref{fig:c22Avsdeta} shows the three-particle correlator for the second
harmonic as a function of $\Delta\eta = \eta_1-\eta_3$.  The left panel shows
the result where $\langle q_3 \rangle$ is subtracted, meaning the same average
charge is subtracted for each charge.
This correlator is likely proportional to $\frac{v_n^2B(\Delta\eta)}{dN/d\eta}$,
where $B(\Delta\eta)$ is the charge balance function.  See~\cite{refScott} for
a treatment of LCC and the charge balance function as related to searches for
local P-violating effects and~\cite{refSergei} for a discussion of those
effects on this particular observable.
Contrariwise, the right panel shows the charge-dependent
subtraction of $\langle q_3^{\pm} \rangle$, meaning the correlation between the
the charge of particle 1 
($q_1$) and $q_3$ is taken into account.  One sees a substantial reduction in the
effect in terms of both strength and in range, i.e. the length of the correlation
in $\Delta\eta$.
From this we may conclude that this subtraction removes some amount of the
LCC contribution, though a detailed theoretical treatment is needed.

\begin{figure}[h!]
\begin{center}
\includegraphics*[width=8.0cm]{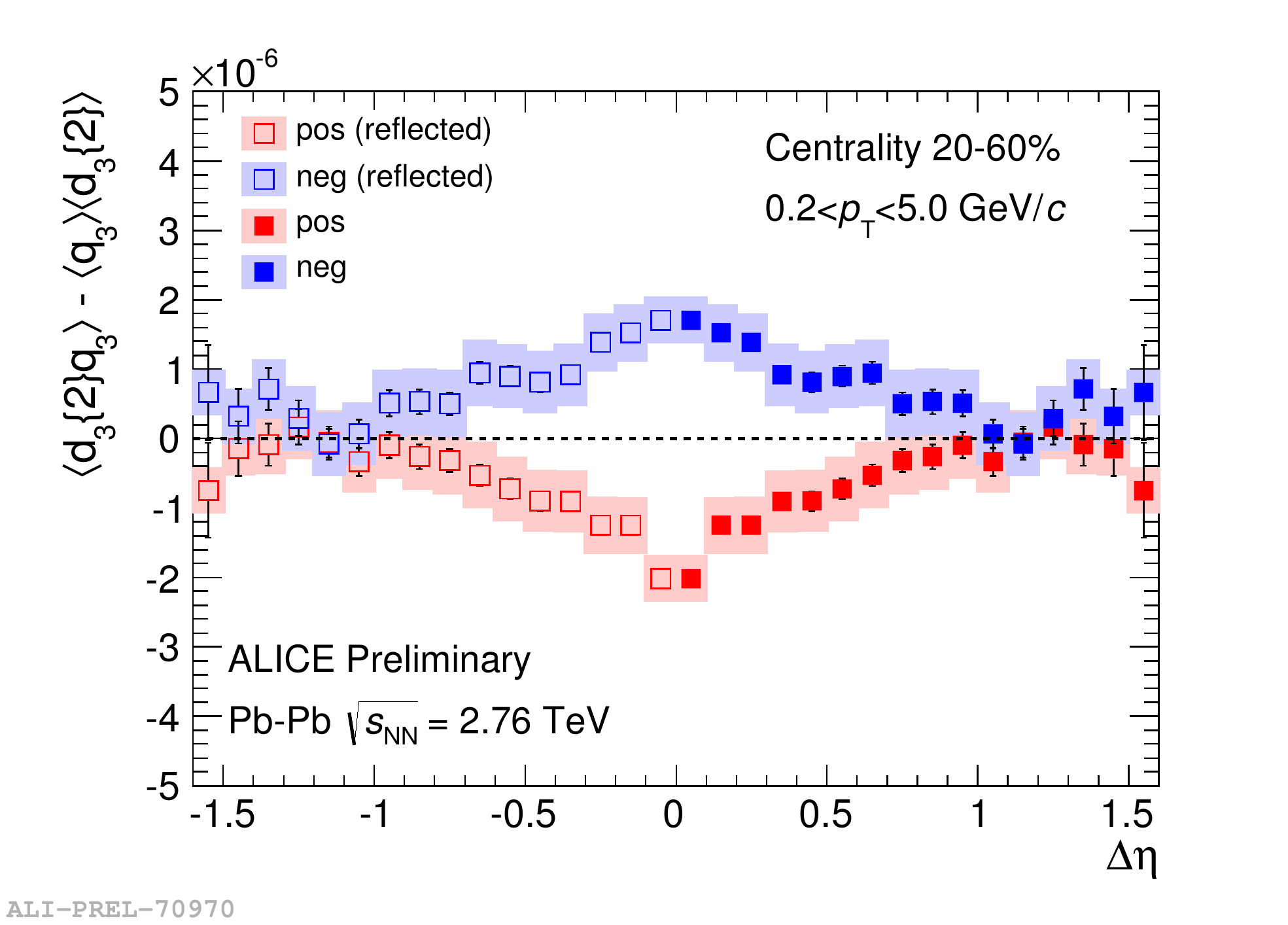}
\includegraphics*[width=8.0cm]{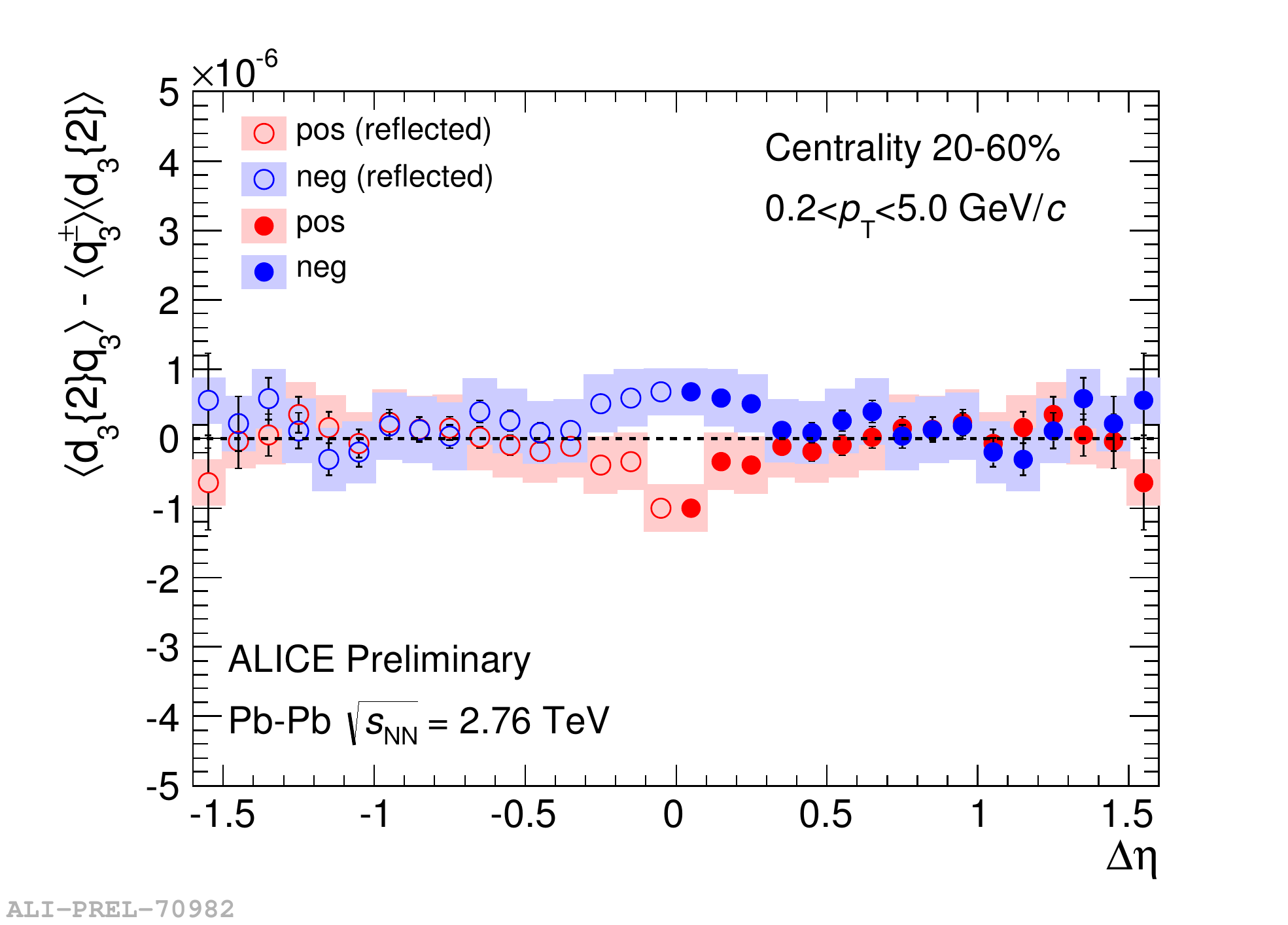}
\caption{Three particle correlator for the third harmonic.
The left panel shows the results with charge independent subtraction
and the right panel shows the results with charge dependent subtraction.}
\label{fig:c32Avsdeta}
\end{center}
\end{figure}

Figure~\ref{fig:c32Avsdeta} shows the three-particle correlator for the third
harmonic as a function of $\Delta\eta$.  As for the previous figure, the left
panel shows the charge independent subtraction and the right shows the charge
dependent subtraction.  Note that the correlation strength
for the unsubtracted is rather strong, whereas the charge dependent subtraction
removes the correlation almost entirely.

\section{Conclusions}

A novel three-particle correlator is employed to search for the CMW.
The results for the second and third harmonic were shown as a function
of centrality and $\Delta\eta$.
Although LCC is thought to be a major background to these measurements,
the charge dependent subtraction can reduce this effect.
Even after the reduction some charge dependent signal is still observed,
which may indicate that some LCC effect remains,
or which may be due to P-violating effects like the CMW,
or some combination of factors.
Further input
from theory is needed to give detailed constraints on the magnitude and range
of LCC vs CMW correlations.








\end{document}